\documentclass[journal]{IEEEtran}
\usepackage{graphicx}
\usepackage[cmex10]{amsmath}
\usepackage{amssymb}
\usepackage{cite}
\begin{document}
% paper title can use linebreaks \\ within to get better formatting as desired
\title{Fourier Response of a Memristor: Generation of High Harmonics with Increasing Weights}
\author{Yogesh~N.~Joglekar~and~Natalia~Meijome
\thanks{Manuscript received \today.}
\thanks{The authors are with the Department of Physics, Indiana University Purdue University Indianapolis (IUPUI), Indianapolis, IN, 46202 USA (email: yojoglek@iupui.edu). This work was supported by the National Science Foundation Grant DMR-1054020 and the Center for Research and Learning, IUPUI.}}% 

%\markboth{Journal of \LaTeX\ Class Files,~Vol.~6, No.~1, January~2007}%
%{Shell \MakeLowercase{\textit{et al.}}: Bare Demo of IEEEtran.cls for Journals}
\maketitle
%----------------------------------------------------------%

\begin{abstract}
%\boldmath
We investigate the Fourier transform of the current through a memristor when the applied-voltage frequency is smaller than the characteristic memristor frequency, and the memristor shows hysteresis in the current-voltage plane. We find that when the hysteresis curve is ``smooth'', the current Fourier transform has weights at odd and even harmonics that decay rapidly and monotonically with the order of the harmonic;  when the hysteresis curve is ``sharp'', the Fourier transform of the current is significantly broader, with non-monotonic weights at high harmonics. We present a simple model which shows that this qualitative change in the Fourier spectrum is solely driven by the saturation of memristance during a voltage cycle, and not independently by various system parameters such as applied or memristor frequencies, and the non-linear dopant drift.
\end{abstract}

% \fi
\IEEEpeerreviewmaketitle
%----------------------------------------------------------%

\section{Introduction}
\label{sec:intro}
\IEEEPARstart{S}{ince} the first experimental demonstration in 2008 by the Williams group at Hewlett-Packard~\cite{strukov}, over the past four years, memristors~\cite{chua1} and memristive systems~\cite{chua2} have been extensively investigated on theoretical and experimental fronts~\cite{review}. This ongoing research is spread across diverse areas such as modeling memristive systems~\cite{drift1,drift2,drift3,drift4,drift5} (including memcapacitors and meminductors)~\cite{memcap1,memcap2}, developing memristor emulators~\cite{em1,em2}, the potential use of memristors for high-density non-volatile memory~\cite{mem1,mem2,mem3}, and the development of artifical neural networks~\cite{neuro0,neuro1,neuro2,neuro3}. The existence of memristive systems (or a memristor) was postulated nearly four decades ago based on symmetry arguments~\cite{chua1}, and many examples of their realization, including the charge transport across ion channels~\cite{hh}, were discussed~\cite{chua2}. However, the first simple, physical model that leads to a charge- (or flux-) dependent resistance was developed by the Williams group~\cite{strukov}; although this model is primarily applicable to a semiconductor thin-film device, it has become a starting point for most of the recent theoretical studies. The two unusal and remarkable properties of a memristor are its hysteretic behavior in the current-voltage characteristics $i(v)$ and the qualitative change in its hysteresis loop structure that takes place when the frequency of the voltage source is much smaller than the characteristic frequency of the memristor in the presence of the applied voltage.\footnote{We use the terms ``memristor'' and ``memristive system'' interchangeably.}  

Historically, hysteretic behavior is associated with magnetic materials where it has been extensively explored both theoretically and experimentally. Since the dependence of the magnetic induction $B$ on the field intensity $H$ is too complex for an analytical treatment, based on approximate models~\cite{stoner}, Fourier analysis has been used to study the dependence of magnetization $M(H)$ and $B(H)$ on a sinusoidally varying field intensity~\cite{f1,f2,f3}. A corresponding Fourier analysis of the response of basic electrical circuit elements to a sinusoidally varying voltage is trivial; it is nonzero only at the frequency of the applied voltage. Similarly, Fourier analysis of non-linear, passive elements is straightforward. Memristors, with their hysteretic properties and a simple, analytical model of their behavior, are ideally suited for Fourier analysis, although such properties are barely explored~\cite{shg}. 

In this paper, we present numerical and analytical results for the Fourier response of the current $i(t)$ through a single memristor with an applied sinusoidal voltage $v(t)=v_0\sin(\omega_{a}t)$. Our two salient results are as follows: (i) A smooth current hysteresis leads to the {\it presence of both even and odd harmonics} of the applied voltage frequency $\omega_a$, and the weight of these harmonics decreases monotonically with their degree. (ii) When the current hysteresis changes to a sharp, switching behavior, the Fourier transform of the current shows a broad, non-monotonic structure; in particular, {\it it has maximum weight at the second harmonic} over a large range of parameters. 

The plan of the paper is as follows. In Sec.~\ref{sec:memmodel}, we describe the memristor model that takes into account the nonlinear dopant drift. We present typical numerical results for the Fourier analysis of the current and its correlations with the system parameters in Sec.~\ref{sec:num}. In Sec.~\ref{sec:theory}, we present a simple analytical model of the hysteretic characteristics and show that this model qualitatively explains the numerical results. We conclude the paper with a brief discussion in Sec.~\ref{sec:disc}.   
%----------------------------------------------------------%

\section{Description of the memristor model}
\label{sec:memmodel}

We use the model of a memristor as two resistors in series~\cite{strukov}: its effective resistance is given by $M(w)=(w/D) R_{\mathrm{on}} + (1-w/D)R_{\mathrm{off}}$, where $w(t)/D=x(t)$ is the fractional width of the doped region with $0\leq x\leq 1$, $D$ is total width of the thin-film device, $R_{\mathrm{on}}$ is the resistance of the memristor if the entire device is doped, and $R_{\mathrm{off}}\gg R_{\mathrm{on}}$ is the resistance of the undoped device. The motion of the boundary between the doped and the undoped regions is approximated by the drift velocity of (oxygen vacancy) dopants, and is characterized by the dopant mobility $\mu_D$. The equations describing the time-evolution of a circuit with a single memristor and a voltage source are given by 
\begin{eqnarray}
\label{eq:iv}
i(t) & = & G(w(t))v(t),\\
\label{eq:drift}
\frac{dw}{dt}& = &\eta\frac{\mu_D R_{\mathrm{on}}}{D}i(t)F\left(\frac{w(t)}{D}\right),
\end{eqnarray}
where the polarity $\eta=\pm 1$ denotes that the doped region increases (decreases) with an applied positive voltage, and the dimensionless window function $F(x)$ is used to model the non-linear drift of dopants near the thin-film boundaries, $x(t)\sim 0$ or $x(t)\sim 1$~\cite{drift2,drift3,drift4,drift5}. The memductance $G(w(t))$ is  
\begin{equation}
\label{eq:mem}
G(w(t))=\frac{1}{M(w(t))}=\frac{G_{\mathrm{on}}G_{\mathrm{off}}}{G_{\mathrm{on}}-x(t)\Delta G},
\end{equation}
with $G_{\mathrm{on}}=R^{-1}_{\mathrm{on}}, G_{\mathrm{off}}=R^{-1}_{\mathrm{off}}$, and $\Delta G= G_{\mathrm{on}}-G_{\mathrm{off}}\gg G_\mathrm{off}$. In the following numerical calculations, we use $R_{\mathrm{on}}$ as the unit of resistance, $i_0=v_0/R_\mathrm{on}$ as the unit of current, and note that the characteristic frequency of the memristor is given by $\omega_{m}=2\pi\mu_D v_0/D^2$.  To simplify the numerical analysis we use the frequency of the applied voltage $\omega_a$ as the unit of frequency (and its period $T_a=2\pi/\omega_a$ as the unit of time), instead of the conventional practice where the memristor frequency sets the scale~\cite{strukov,drift2,drift5}. The hysteretic effects for different values of $\Omega=\omega_a/\omega_m\lesssim 1$ are investigated by varying $v_0$ and thus $\omega_m$. Typical memristor parameters~\cite{strukov} are $R_\mathrm{on}\sim 10^3$ Ohm, $D\sim10$ nm, $\mu_D\sim 10^{-14}$ cm$^2$V$^{-1}$s$^{-1}$, and thus imply that the memristor frequency is $\omega_m\sim 50$ KHz and the current is $i_0=1$ mA for $v_0\sim 1$ V.

We use the following one-parameter family of window functions
\begin{equation}
\label{eq:window}
F_p(x)=1-(2x-1)^{2p}
\end{equation}
which ensures that the dopant drift is symmetrically suppressed as $x\rightarrow 0,1$~\cite{drift2}. Note that Eq.(\ref{eq:window}) can be generalized to non-integer values of the parameter 
$p\geq 1$ by using the absolute value of the argument, $|2x-1|$. Since the window function vanishes at the boundaries, $F(0)=0=F(1)$, it has two fixed points in the time-evolution of the fractional width; however, this is not of physical concern for the following reason. Suppose the fractional width changes from an initial value $x_1=1-\delta_1$ to a new value $x_2=1-\delta_2$ after the passage of charge $Q$ where $\delta_2\ll\delta_1\ll 1$. This process describes the approach to the fixed point $x=1$. The charge can be estimated from Eq.(\ref{eq:drift}) by approximating the window function near the boundary, $F(x=1-\delta)\approx 4p\delta$, 
\begin{equation}
\label{eq:charge}
Q= \frac{\eta D^2}{4p\mu_D R_{\mathrm{on}}}\ln\left(\frac{\delta_1}{\delta_2}\right).
\end{equation}
Eq.(\ref{eq:charge}) shows that the charge $Q$ required to reach the fixed point, $\delta_2=0$, diverges. Thus, starting from an initial width $x\in (0,1)$ it is impossible to come across the ``terminal state problem''~\cite{drift3,drift4} in a finite amount of time. An identical argument applies to a memristor where the doped region decreases with time, thus approaching the $x=0$ fixed point. Therefore, in the following, we use the phrase ``saturation of the fractional width'' to denote fractional width $x(t)$ arbitrarily close to one (zero) without being exactly equal to one (zero). 
 
%----------------------------------------------------------%

\section{Fourier analysis: numerical results}
\label{sec:num}

We calculate the (discrete) Fourier transform (FT) of the memristor current $i(t)$ that, in turn, is obtained by numerically solving Eqs.(\ref{eq:iv})-(\ref{eq:drift}) over many periods $T_a=2\pi/\omega_a$ of the applied sinusoidal voltage to ensure stability. The discrete FT of the current is given by 
\begin{equation}
{\mathcal I}(\nu_k)=\frac{1}{N}\sum_{j=1}^{N} e^{i\nu_k t_j} i(t_j)={\mathcal I}^{*}(-\nu_k)
\end{equation}
where $N$ is the number of points that approximate the continuous function $i(t)$ over a single period, $t_j=(j-1)\Delta t=(j-1)T_a/N$, and $\nu_k=k\omega_a$ for $k=-N/2,\ldots,N/2-1$ are the $N$ possible discrete frequencies; we verify that our results do not depend upon $N$ and thus are truly representative of the continuum limit, $N\rightarrow0, \Delta t\rightarrow 0$. 

% Hysteresis iv plane.
\begin{figure}[htb]
\centering
\includegraphics[width=1.1\columnwidth]{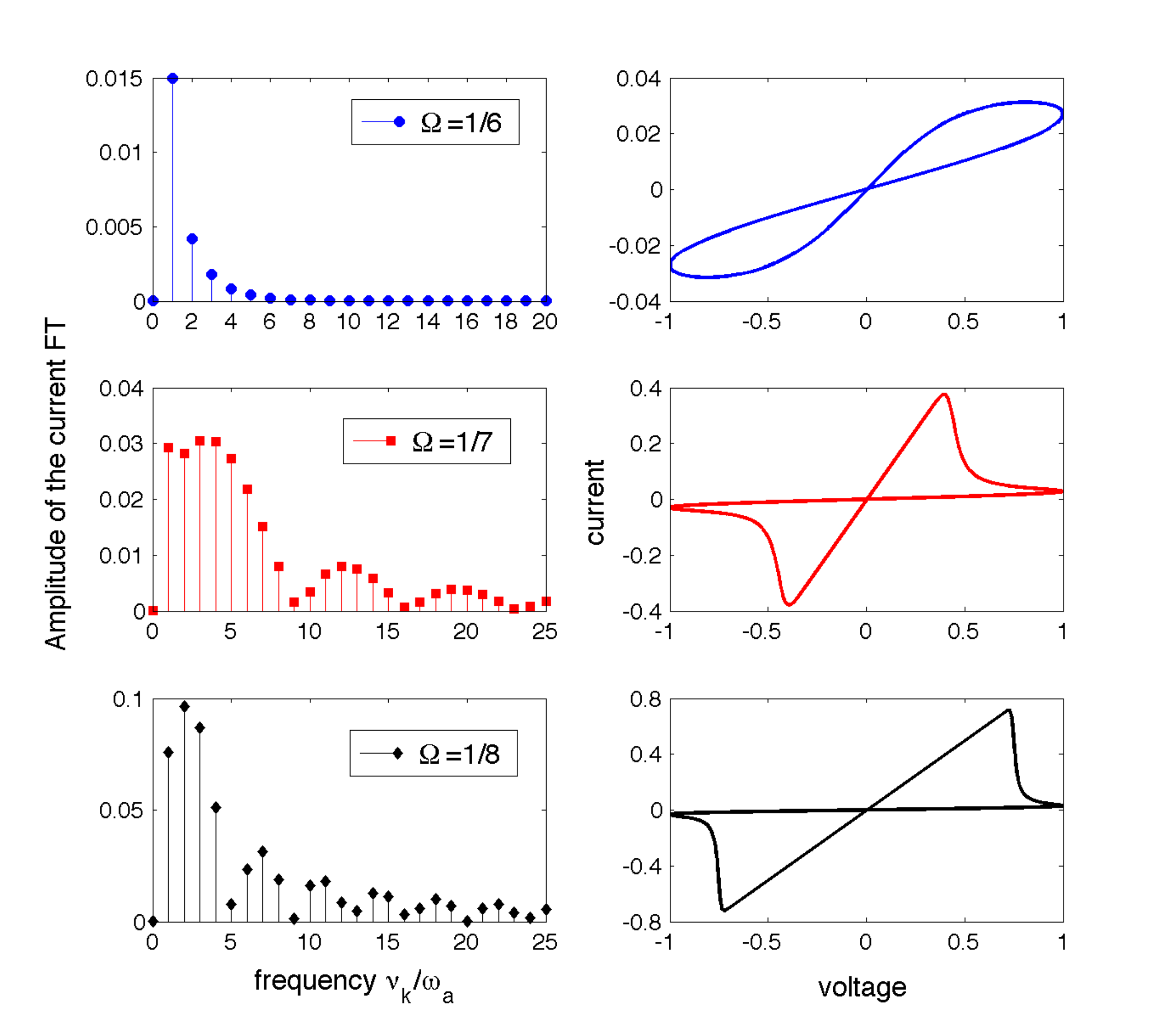}
\caption{The left-hand column shows the amplitude of current FT, $|{\mathcal I}(\nu_k)|$ as a function of frequency $\nu_k/\omega_a$, and the right-hand column shows corresponding time-domain $i(v)$ curves. As the ratio of applied voltage frequency to the memristor frequency $\Omega=\omega_a/\omega_m$ decreases, the hysteresis changes from smooth (top row) to sharp (center and bottom row), and the current FT changes from monotonic decay (blue circles) to a broad, bumpy structure (red squares, black diamonds) with a maximum weight at the second harmonic (black diamonds).}
\label{fig:ivhyst}
\end{figure}
Figure~\ref{fig:ivhyst} shows typical results of such an analysis. The left-hand column shows the amplitude of the FT of current, $|{\mathcal I}(\nu_k)|=|{\mathcal I}(-\nu_k)|$ and the right-hand column shows corresponding time-domain $i(v)$ curves plotted in the units of $i_0$ and $v_0$. Note that since the time-integral of the current over a single period is zero, the spectral weight at $k=0$ vanishes in all cases. These results are obtained with $R_\mathrm{off}/R_\mathrm{on}=100$, $x(t=0)=0.5$, $\eta=+1$, and window function parameter $p=5$. When $\Omega=\omega_a/\omega_m\gtrsim 1$, the memristor behaves essentially as an ohmic resistor with constant value and the current FT is nonzero only at $\nu_k=\pm\omega_a$. 

As $\Omega$ decreases (top row, $\Omega=1/6$) a smooth, pinched hysteresis loop develops in the $i(v)$ plane. The maximum current is $i/i_0\sim R_\mathrm{on}/R_\mathrm{init}\sim 0.02$ where $R_\mathrm{init}=0.5 R_\mathrm{on} +(1-0.5)R_\mathrm{off}$ is the initial resistance of the memristor. The corresponding current FT (filled blue circles) shows finite, monotonically and rapidly decreasing weights $|{\mathcal I}(\nu_k)|$ at harmonics $\nu_k=k\omega_a$ for $1\leq k\lesssim 10$. These even and odd higher harmonics are generically present in the regime where the memristor shows hysteresis. When $\Omega=1/7$ (center row), the $i(v)$ hysteresis loop now has a butterfly shape,  and the maximum current is now determined by the significantly smaller resistance value $R\sim R_\mathrm{on}\ll R_\mathrm{init}$ that the memristor reaches when $x(t)$ approaches one. The corresponding current FT (filled red squares) shows a broad structure with appreciable, non-monotonic weights at high harmonics $\nu_k$ with $k\lesssim 25$. The bottom row shows that for $\Omega=1/8$, the $i(v)$ curve exhibits a clear switching behavior from a state with high resistance $R_\mathrm{init}$ to a state with very low resistance $R\sim R_\mathrm{on}$ accompanied by the saturation of the fractional width $x(t)$. The Fourier transform of the current (filled black diamonds) shows that a markedly higher weight at the second harmonic develops compared to the weight at the first harmonic. Note that since the maximum current through the memristor changes by an order of magnitude or more as $\Omega$ decreases, the ``small spectral weight'' at high harmonics for $\Omega=1/8$ (bottom row) is, in absolute terms, comparable to the maximum spectral weight of the first harmonic for $\Omega=1/6$ (top row); if the latter is experimentally measurable, so will be the former. 

Thus, the Fourier analysis of the memristor current shows two main features. The first is that a smooth hysteresis curve is accompanied by a current spectrum with monotonically decreasing weights at both even and odd harmonics. The second is that this spectrum changes to a broad, non-monotonic, bumpy  structure with nonzero weights at very high harmonics as the hysteresis curve changes to a butterfly shape characterized by two resistance values. The question is: is it the frequency $\Omega=\omega_a/\omega_m$ that drives this change? 

% Current FT vs. width-time.
\begin{figure}[htb]
\centering
\includegraphics[width=1.1\columnwidth]{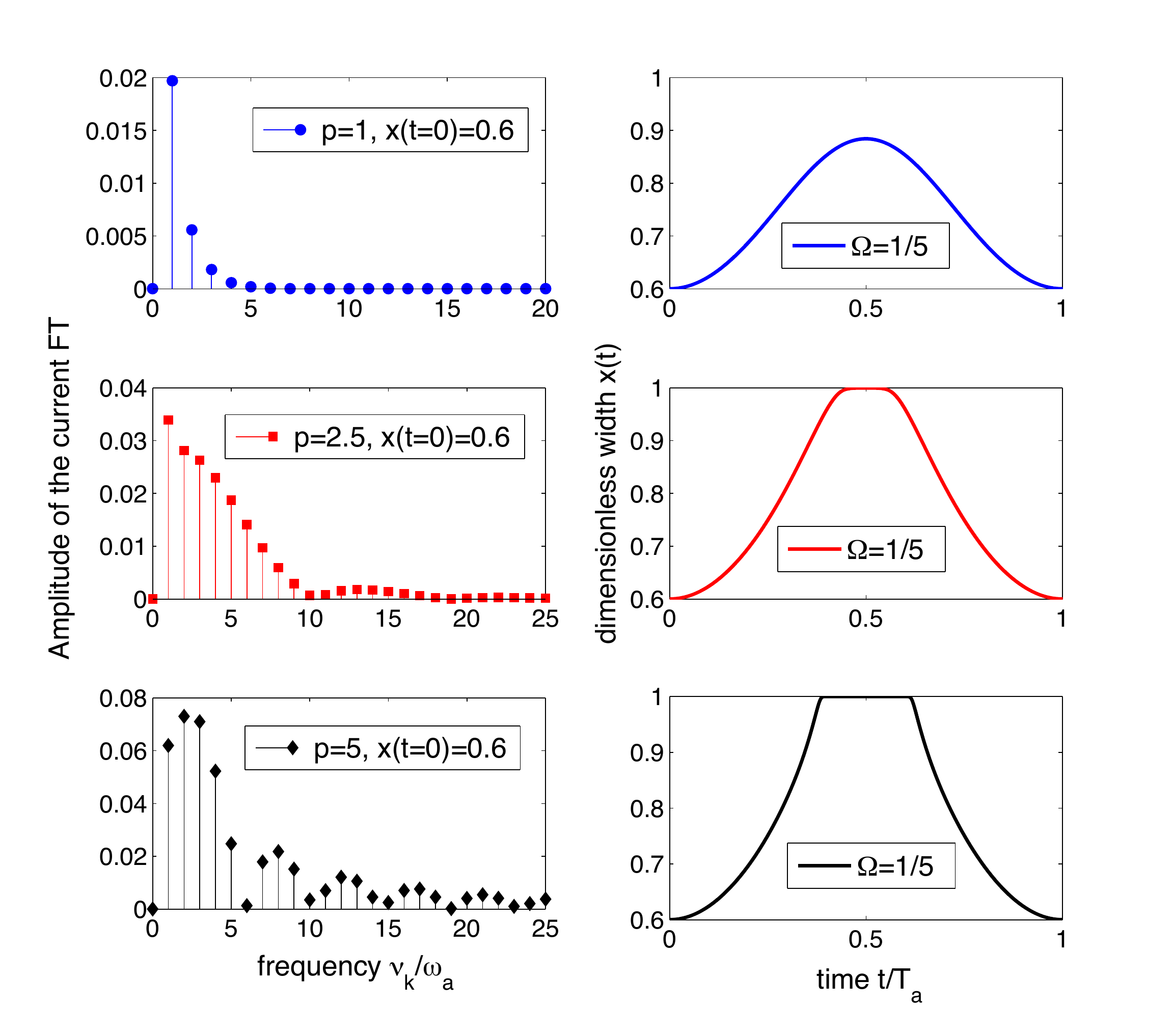}
\caption{The left-hand column shows the amplitude of current FT, $|{\mathcal I}(\nu_k)|$ as a function of frequency $\nu_k/\omega_a$ and the right-hand column shows the time-evolution of the fractional width $x(t)$ over a single period $0\leq t/T_a\leq 1$ for window functions with $p=1$ (top row), $p=2.5$ (center row), and $p=5$ (bottom row). These typical results suggest that the qualitative change in the current FT is triggered by the saturation of the fractional width.}
\label{fig:hystp}
\end{figure}
To investigate the connection, if any, between the broad spectrum of the current FT and the saturation of the fractional width, we obtain $|{\mathcal I}(\nu_k)|$ for different window functions while keeping all other parameters constant. The right-hand column in Fig.~\ref{fig:hystp} shows the time-evolution of $x(t)$ from its initial value $x(t=0)=0.6$ for three values of $p=\{1, 2.5, 5\}$; the left-hand column shows corresponding amplitude of the current FT. When $p=1$ the highly non-linear window function reduces the dopant drift velocity and therefore the fractional width reaches a maximum $x(t=T_a/2)\sim 0.9$ that is well below saturation (top row); the corresponding $|{\mathcal I}(\nu_k)|$ shows monotonically decreasing weights at harmonics $\nu_k=k\omega_a$ for $2\leq k\lesssim 10$ (filled blue circles). For $p=2.5$ the dopant drift is higher, the fractional width barely reaches saturation at $t=T_a/2$ (center row), and the current FT now shows a broad, non-monotonic, bumpy structure (filled red squares). The bottom row shows that when $p=5$, the dopant drift is essentially constant, the dimensionless width saturates to one for a significant fraction of the period $T_a$, and the current spectrum now shows maximum weight at the second harmonic (filled black diamonds). We emphasize that these results are for a single value of the memristor frequency, $\Omega=1/5$. 

These numerical results strongly suggest that saturation of the fractional width is instrumental to the qualitative change in the current spectrum and the emergence of higher-weight second harmonic. They show, in particular, that neither the frequency $\Omega$ nor the dopant drift dynamics, characterized by the window function index $p$, solely drive the emergence of higher harmonics with increasing weights. 

%----------------------------------------------------------%

\section{Fourier analysis: theoretical model}
\label{sec:theory}

Armed with these insights, we now present a theoretical model that qualitatively explains the properties of current Fourier transform in terms of time evolution of the memductance. In general, the Fourier transform of current $i(t)$ is given by the convolution, 
\begin{equation}
\label{eq:convo}
{\mathcal I}(\nu_k)=\sum_{j=-N/2}^{N/2-1}{\mathcal G}(\nu_k-\nu_j){\mathcal V}(\nu_j),
\end{equation}
where ${\mathcal G}$ and ${\mathcal V}$ denotes Fourier transforms of the memductance $G(t)$ and the applied voltage respectively. For a sinusoidal voltage, Eq.(\ref{eq:convo}) becomes 
\begin{equation}
\label{eq:cv2}
{\mathcal I}(\nu_k)=\frac{v_0}{2i}\left[{\mathcal G}(\nu_{k-1})-{\mathcal G}(\nu_{k+1})\right],
\end{equation}
where we have used $\nu_k\mp\omega_a=\nu_{k\mp 1}$. Thus, an analytical model for the memductance FT ${\mathcal G}(\nu_k)$ can shed light on the evolution of the current Fourier transform. 

We start with the case when the fractional width $x(t)$ does not saturate (top-right panel in Fig.~\ref{fig:hystp}). Then the current can be expressed as sum of two, low-order, odd polynomials in voltage
\begin{equation}
\label{eq:pq}
i(t)=P(v(t))\theta(v'(t))+Q(v(t))\theta(-v'(t)),
\end{equation}
where $\theta(x)$ is the Heaviside function, and the sign of $v'(t)=dv/dt$ denotes whether the voltage is increasing with time or decreasing. For example, the hysteresis for $\Omega=1/6$ in Fig.~\ref{fig:ivhyst} can be modeled by using $P(x)=a_1 x + a_3 x^3$ and $Q(x)= b_1 x - b_3 x^3$ where $a_i, b_i>0$ and $a_1=G_\mathrm{init}< b_1$, whereas when $\Omega\sim 1$, $P(x)=a_1x=Q(x)$. Although the polynomials are odd functions of time, due to the presence of the Heaviside function, the Fourier transform of Eq.(\ref{eq:pq}) contains both even and odd harmonics. We note that although such approximation works well for the most part, for any degree polynomial, it cannot capture the diverging slope of the $i(v)$ curve that occurs at the voltage extrema, $v=\pm v_0$. For a power-law current with a single exponent, $i=A_n v^{n}$, and sinusoidal voltage, the Fourier transform is given by 
\begin{equation}
\label{eq:power}
{\mathcal I}(\nu_k)=A_n \frac{v_0^n}{(2i)^n}{n\choose \frac{n-k}{2}}.
\end{equation}
The binomial coefficient, and therefore the spectral weight $|{\mathcal I}(\nu_k)|$, decreases monotonically with the order of the harmonic $1\leq k\leq n$; note that ${\mathcal I}(\nu_k)=0$ 
for $k>n$. It follows from Stirling's approximation that the current FT will decay rapidly as $\exp(-k^2/n)$ for large $k$. This exponential nature of the decay does not depend upon the power-law exponent $n$, and is applicable to a general polynomial $i=P(v)$. Thus, the two-polynomial model qualitatively explains the monotonically decreasing weight at even and odd harmonics for the current FT. 

% Step function behavior of memductance.
\begin{figure}[htb]
\centering
\includegraphics[width=1\columnwidth]{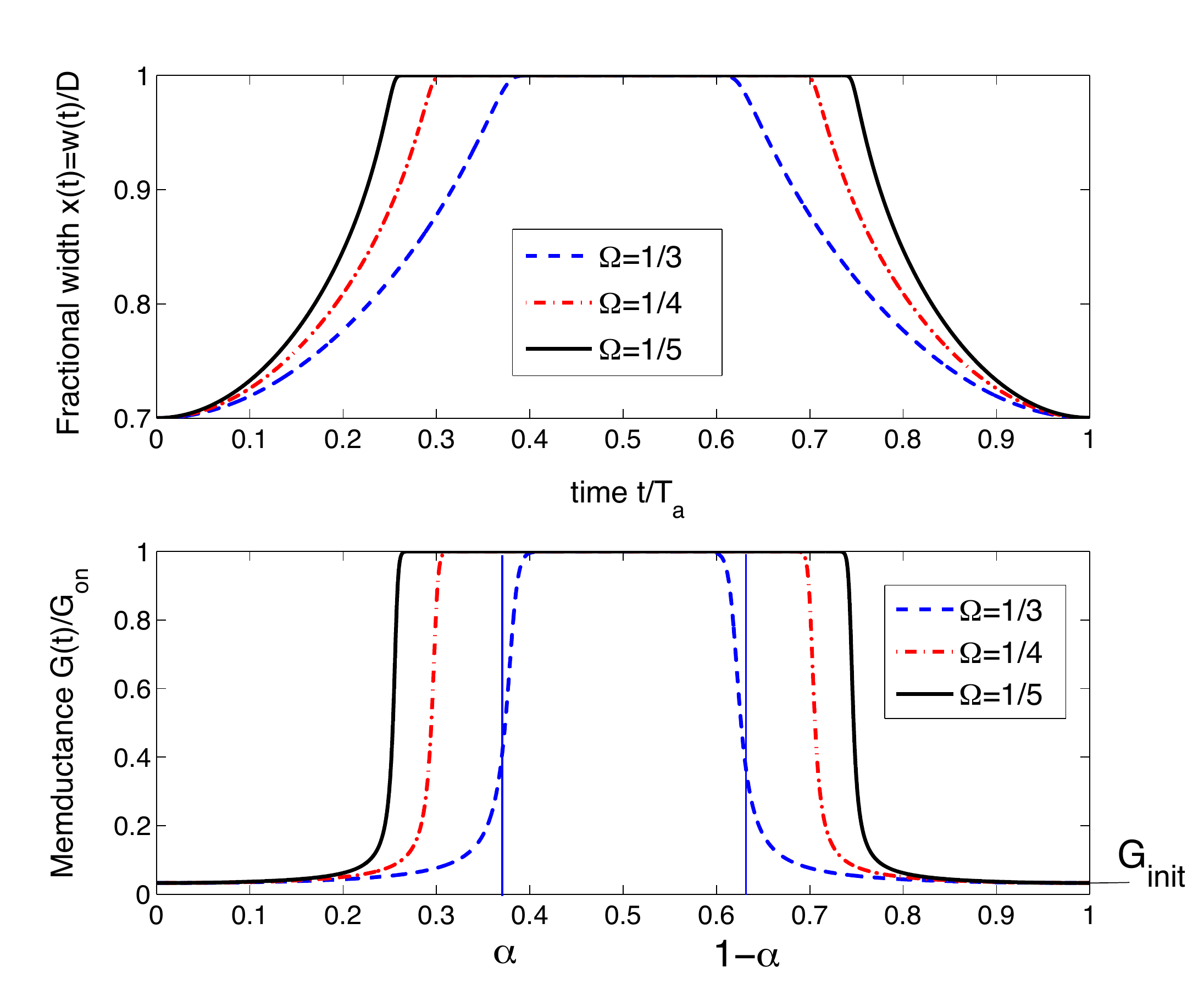}
\caption{Time evolution of the fractional width $x(t)$ (top panel) and the memductance $G(t)/G_\mathrm{on}$ (bottom panel) over a single period $0\leq t/T_a\leq 1$. The system parameters are $R_\mathrm{off}/R_\mathrm{on}=100$, $p=5$, $\eta=1$, $x(t=0)=0.7$, which implies that $R_\mathrm{init}=30.7 R_\mathrm{on}$. The results show that although the fractional width slowly saturates to one over time (top panel), $G(t)$ increases sharply from $G_\mathrm{init}=0.033 G_\mathrm{on}$ to $G_\mathrm{on}$ at time $t/T_a=\alpha$ (bottom panel).}
\label{fig:gwvstime}
\end{figure}
When the fractional width $x(t)$ saturates at low frequencies, the memductance $G(t)$ increases from its initial value $G_\mathrm{init}=R^{-1}_{\mathrm{init}}\ll G_\mathrm{on}$ to a maximum value of $G_\mathrm{on}$. Figure~\ref{fig:gwvstime} shows the typical time evolution of the fractional width $x(t)$ (top panel) and the corresponding memductance $G(t)$ (bottom panel) as a function of $\Omega$. In general $G(t)$ can be approximated by a piecewise constant function,
\begin{equation}
\label{eq:step}
G(t)=\left\{
\begin{array}{cc}
G_\mathrm{init} & 0\leq t/T_a < \alpha, \\
G_\mathrm{on} & \alpha \leq t/T_a< 1-\alpha,\\
G_\mathrm{init} & 1-\alpha \leq t/T_\alpha\leq 1,\\
\end{array}
\right.
\end{equation}
where $0<\alpha<1/2$ is determined by the system parameters; for example, $\alpha\approx 0.37$ when $\Omega=1/3$ and monotonically decreases with $\Omega$. Its Fourier transform is given by
\begin{equation}
\label{eq:gft}
{\mathcal G}(\nu_k)=G_\mathrm{on}\delta_{k,0} -2\alpha(G_\mathrm{on}-G_\mathrm{init})\frac{\sin(2\pi\alpha k)}{2\pi\alpha k}.
\end{equation}
The presence of the sinc-function in ${\mathcal G}(\nu_k)$ and the fact that $G_\mathrm{on}\gg G_\mathrm{init}$ imply that memductance spectral weight $|{\mathcal G}(\nu_k)|$ is non-monotonic and its higher harmonics decay slowly as $1/k$ for large $k$. Then it follows from Eq.(\ref{eq:cv2}) that the broad, bumpy structure of the current FT arises from the same source. 

%----------------------------------------------------------%

\section{conclusions}
\label{sec:disc}
In this paper, we have presented numerical results for the Fourier response of the current through a single memristor in the presence of a sinusoidal voltage. With simple models, we have shown that the qualitative change in the current FT can be traced to the qualitative change in the hysteresis loop in the $i(v)$ plane. The results presented here are for a positive polarity, $\eta=+1$, where the fractional width saturates to one. 

When $\eta=-1$, the memductance of the device {\it decreases} from $G_\mathrm{init}$ to $G_\mathrm{off}\lesssim G_\mathrm{init}$. Thus, to obtain a ratio of the initial and saturation memductance values comparable to the positive polarity case, we will need an initial fractional width $x(t)\sim 1$. This, in turn, implies that the frequency $\Omega$ required for width saturation down to zero is much smaller than the corresponding values in the positive polarity case. Therefore, non-monotonic current spectrum and higher-weight second harmonic response~\cite{shg} in negative polarity memristors require significantly different conditions. 

Our approach can be used to analytically investigate other phenomena, such as power in a load resistor $R_L$ in series with a memristor, $P(t)=i^2(t)R_L$~\cite{shg}. Since $R_L$ is constant, the form the current-voltage hysteresis and the subsequent results remain the same. Since ${\mathcal P}(\nu_k)=\sum_j{\mathcal  I}(\nu_k-\nu_j){\mathcal I}(\nu_j)$ where ${\mathcal P}(\nu_k)$ is the power FT, our results can be used to understand its behavior~\cite{shg}. 

%\hfill January 11, 2007
%\IEEEpubidadjcol

% An example of a double column floating figure using two subfigures.
% (The subfig.sty package must be loaded for this to work.)
% The subfigure \label commands are set within each subfloat command, the
% \label for the overall figure must come after \caption.
% \hfil must be used as a separator to get equal spacing.
% The subfigure.sty package works much the same way, except \subfigure is
% used instead of \subfloat.
%
%POTENTIAL ERROR: FIX ON SHELLs PAGE
%\begin{figure*}[!t]
%\centerline{\subfloat[Case I]\includegraphics[width=2.5in]{subfigcase1}%
%\hfil
%\subfloat[Case II]{\includegraphics[width=2.5in]{subfigcase2}%
%\label{fig_second_case}}}
%\caption{Simulation results}
%\label{fig_sim}
%\end{figure*}

\section*{Acknowledgment}
The authors would like to thank Profs. Ricardo Decca and Gautam Vemuri stimulating discussions. 


%\begin{IEEEbiography}{Yogesh N. Joglekar}
%Yogesh N. Joglekar received the integrated M.Sc. degree in physics from Indian Institute of Technology, Kanpur, India in 1996 and the Ph.D. degree in physics from Indiana University, Bloomington in 2001. \\ Since 2006, he has been with the Department of Physics, Indiana University - Purdue University Indianapolis, where he is an Associate Professor. His research interests include graphene, excitonic condensates, PT symmetry, and memristrive systems.
%\end{IEEEbiography}
%\begin{IEEEbiography}{Natalia Meijome}
%Natalia Meijome received the B.S. degree in physics from Indiana University- Purdue University Indianapolis (IUPUI), Indianapolis in 2012. She will join the Ph.D. program in neuroscience at Purdue University, West Lafayette this Fall. \\ Her research interests include memristive systems and neural networks.
%\end{IEEEbiography}

%----------------------------------------------------------%
\end{document}